\documentclass[a4paper]{JHEP3} 
\usepackage{amsfonts}
\usepackage{amsmath,amssymb}

\def\p{\partial}
\def\half{\frac{1}{2}}
\def\real{\mathbb{R}}
\def\complex{\mathbb{C}}
\def\act{\triangleright}
\def\be{\begin{equation}}
\def\ee{\end{equation}}
\def\bea{\begin{eqnarray}}
\def\eea{\end{eqnarray}}
\def\and{\quad{\rm and}\quad}

\def\CA{{\cal A}}   \def\CD{{\cal D}} 
 \def\CF{{\cal F}}  \def\CH{{\cal H}} 
    
 \def\CN{{\cal N}} \def\CO{{\cal O}} \def\CP{{\cal P}}
 \def\CR{{\cal R}}

\title{Twist Quantization of String and $B$ Field Background}

\author{Tsuguhiko Asakawa${}^{a}$, Masashi Mori${}^{b}$ 
and Satoshi Watamura${}^{b}$\\
${}^a$Department of Physics, 
Hokkaido University\\
Sapporo, 060-0810, Japan \\
${}^b$Department of Physics,
Graduate School of Science,
Tohoku University,\\
Sendai 980-8578, Japan \\
E-mail: \email{asakawa@particle.sci.hokudai.ac.jp},
\email{morimasa@tuhep.phys.tohoku.ac.jp},
\email{watamura@tuhep.phys.tohoku.ac.jp}}

\preprint{EPHOU-08-005\\Preprint TU-827} 

\abstract{%
In a previous paper, we investigated the Hopf algebra structure in string theory 
and gave a unified formulation of the quantization of the string 
and the space-time symmetry.
In this paper, this formulation is applied to the case with a nonzero
 $B$-field background, and the twist of the Poincar\'e symmetry is studied.
The Drinfeld twist accompanied by the $B$-field background 
gives an alternative quantization scheme, 
which requires a new normal ordering.
In order to obtain a physical interpretation 
of this twisted Hopf algebra structure,
we propose a method to decompose the twist into two successive twists
and we give two different possibilities of decomposition.
The first is a natural decomposition from the viewpoint of the twist quantization, leading to a new type of twisted Poincar\'e symmetry.
The second decomposition reveals the relation of our formulation
 to the twisted Poincar\'e symmetry on the Moyal type noncommutative 
space.}

\begin{document}

\section{Introduction}

The connection between string theory and noncommutative geometry 
is an important subject and has been 
developed especially in the case of $D$-branes in the NSNS $B$-field background
(for a review see Refs.\cite{Douglas:2001ba,Szabo:2001kg}).
In a background of constant $B$ field, the effective theory
on the $D$-branes in a certain zero-slope limit is 
described very well by the gauge theory on the noncommutative space
\cite{Seiberg:1999vs}, where
all the products are replaced by the Moyal-Weyl product.
The noncommutativity of the space,
\be
[x^\mu \stackrel{*}{,} x^\nu]=i\theta^{\mu\nu}, 
\ee
originates from the operator product expansion of open string vertex operators 
\cite{Schomerus:1999ug}, 
or equivalently, from the
canonical quantization of open strings in a $B$-field background 
\cite{Chu:1998qz}.
This connection drew the interests towards phenomena like 
UV/IR mixing, noncommutative solitons, 
open Wilson lines, etc., however, the explicit 
breaking of the Lorentz symmetry 
was a bottleneck.

It is known that in a wide class of noncommutative spaces, 
Hopf algebras are considered to be the underlying symmetries 
rather than just groups or Lie algebras
\cite{2000fqgt.book.....M,majid-1999-205,SitarzOctober2001}.
When we consider noncommutative spaces as deformations of commutative spaces,
such Hopf algebras are obtained by a Drinfeld twist of the Hopf algebra of
the corresponding symmetry group \cite{Drinfeld:1990qh} of the underlying 
undeformed spaces.
Therefore, it is natural to expect 
that there is also a Hopf algebra structure
corresponding to the deformed symmetry on the Moyal-Weyl space.
In fact, such twisted Hopf algebra structures were studied and 
the twist element corresponding to the Moyal-Weyl product was discovered
in Refs.\cite{Oeckl:2000eg,Watts:2000mq} for the translation symmetry, 
and in \cite{Oeckl:2000eg} for the Poincar\'e group.
More recently, the twisted Poincar\'e symmetry of the noncommutative field theory 
has been fully realized in 
Refs.\cite{Chaichian:2004za,Koch:2004ud, Aschieri:2005yw}.
There, the Moyal-Weyl product is considered as a twisted 
product equipped with the Drinfeld twist of the Poincar\'e-Lie algebra.
In other words, both the noncommutativity and 
the modification of the symmetry are controlled by a single twist.
This approach is then generalized to the twisted version of the diffeomorphism on the 
Moyal-Weyl space \cite{Aschieri:2005yw,Aschieri:2005zs}
(see for example Ref.\cite{Szabo:2006wx} and references therein on these developments).

Considering the string theory, one may raise the question whether
there is a string theory counterpart of 
these twisted Hopf algebra structure, and if so, how it is
 realized in the string theory.
In this respect, the authors of Ref.\cite{AlvarezGaume:2006bn} 
studied a effective theory on D-branes 
in a zero slope limit, where the leading 
brane-induced gravitational effects are visible, 
and are compared with the noncommutative 
gravity action proposed in Ref.\cite{Aschieri:2005yw}. 
The authors of Ref.\cite{AlvarezGaume:2006bn} were led to a negative answer, since
effective gravity in a $B$-field background has more interaction terms than 
that in Ref.\cite{Aschieri:2005yw}.
However, we have to keep in mind that there are always 
ambiguities when comparing two theories only 
on-shell, and thus it is difficult to
capture the essential difference between these two results
by a comparison of the actions.
It is plausible that a direct comparison of the underlying 
symmetries, without recourse to the actions, will lead to a more 
concrete answer. 
The purpose of this paper is to give a suitable framework 
to answer the above questions.

In a previous paper Ref.\cite{AMW}, 
we have investigated the Hopf algebra structure in string worldsheet theory 
in the Minkowski background
and gave a unified formulation of the quantization of the string and the 
spacetime symmetry, by reformulating 
the path-integral quantization of the string as a 
Drinfeld twist at the worldsheet level.
The coboundary relation showed that 
this was equivalent to operators with normal ordering.
By the twist, the space-time diffeomorphism was deformed 
into a twisted Hopf algebra, while the Poincar\'e symmetry was 
kept unchanged.
This result suggests a characterization of the symmetry as follows: 
unbroken symmetries are twist invariant Hopf subalgebras,
while broken symmetries are realized as twisted ones.

Although in Ref.\cite{AMW} only the case of a Minkowski spacetime 
as a background was considered explicitly, the process is applicable 
to a wider class of backgrounds.
In this paper, we apply our formulation to the case
of the presence of 
a nonzero $B$-field, as the most simple example of such a background,
and show that there is also a natural structure of a twisted Hopf algebra,
which depends on the $B$-field.
As a quantization, this twisted Hopf algebra structure corresponds to 
a new normal ordering in the operator formulation, which was studied 
in Refs.\cite{Braga:2004wr,Braga:2006gi,Chakraborty:2006yj,Gangopadhyay:2007ne}.
As a symmetry, it represents a twisted symmetry including space-time 
diffeomorphisms.

To understand the geometrical picture and the structure 
of the twisted symmetry in a $B$-field background,
we formulate a method to decompose the twist 
into a combined operation of two successive twists.
As we will see, we can interpret the first twist as a quantization 
and the second twist as a deformation of the first twisted Hopf algebra.
We give two types of decomposition. 
The first type is natural from the viewpoint of the twist quantization 
and useful for 
comparing the twisted Hopf algebra structure in $B\ne 0$ 
with the corresponding structure in the case $B=0$, treated 
in Ref.\cite{AMW}.
The result indicates the existence of 
a geometrical spacetime description 
different from the commutative spacetime with $B$-field, 
where the effect of the $B$-field is hidden in a twist element.
This is similar to the noncommutative spacetime description 
used in open string theory, however the formulation using the twist
is valid for closed strings as well.
The second type of decomposition is useful when we want to 
compare the Hopf algebra structure in the open string sector 
with the twisted Poincar\'e symmetry\cite{Chaichian:2004za,Koch:2004ud, Aschieri:2005yw} in the Moyal-Weyl noncommutative space.

The paper is organized as follows:
In section 2 we give a brief review of the 
results of our previous paper, Ref.\cite{AMW} 
and explain some notations, which are necessary to proceed the 
analysis here. 
In section 3, applying the formulation 
proposed in Ref.\cite{AMW}
to the case of non-zero $B$-field background,
we present the twist quantization of the string worldsheet theory
and the corresponding twisted symmetry for the case $B\ne 0$.
In section 4, we first give a method to decompose the twist 
discussed in section 3
into two successive twists, 
and discuss its interpretation.
We apply this splitting method of Hopf algebra twist 
in two different ways.
Section 5 is devoted to discussion and conclusion.

\section{Hopf algebra in string theory}\label{sec:A}

In this section, we briefly review the content of Ref.\cite{AMW}
with a slightly generalized form so that the discussions in the 
following section becomes 
more transparent. 

\subsection{Classical Hopf and module algebras}
\label{sec:classical}

Our starting point is the classical Hopf algebra $\CH$ of 
functional derivatives
and its  module algebra $\CA$ of string functionals.
The functionals depend on a target space manifold $\real^d$ 
but are independent of the background such as 
the metric and $B$ field (see Ref.\cite{AMW} for more details).  

The classical string variable $X^\mu(z)\,(\mu=0,\cdots d-1)$ is a set of 
functions defining an embedding $X:\Sigma \rightarrow \real^d$ of a
 worldsheet into a target space.
Let $\CA$ be the space of the complex valued embedding functionals 
$X^\mu(z)$ together with their worldsheet derivatives $\p_a X^\mu(z)$ of the 
form 
\be
I[X]= \int \!d^2 z\ \rho(z) F[X(z)]~ .
\label{def of functional}
\ee
where $F[X(z)] \in C^\infty(\Sigma)$ 
is a component of a pull-back tensor field 
in the target space
and $\rho(z)$ is some weight function (distribution).
In particular, we call 
\be
F[X](z_i)= \int \!d^2 z\ \delta^{(2)}(z-z_i) F[X(z)]~ ,
\label{local functional}
\ee
a local functional at $z_i$, with an additional label $z_i \in \Sigma$.   
We also write it simply as $F[X(z_i)]$ when it is not confusing. 
These functionals (\ref{def of functional}) and 
(\ref{local functional}) 
correspond to an integrated vertex operator 
and to a local vertex operator after quantization, respectively.

We define a multiplication of two functionals 
$m : \CA\otimes \CA \rightarrow \CA$ as
$I_1I_2[X]=I_1[X]I_2[X]$, where the r.h.s. is the multiplication in $\complex$, 
in particular,
$FG[X](z_1,z_2)=F[X(z_1)]G[X(z_2)]$ for two local functionals.
By including all multi-local functionals with countable labels, $\CA$ forms 
an algebra over $\complex$. 
Note that its product is commutative and associative.

Next, let ${\mathfrak X}$ be the space of all functional vector fields of the form
\be
\label{gene vector field}
\xi = \int \! d^2w \, \xi^\mu (w) \frac{\delta}{\delta X^{\mu}(w)},
\ee
and denote its action on $I[X] \in \CA$ as $\xi \act I[X]$.
Here $\xi^\mu (w)$ is a weight function 
(distribution) on the worldsheet including the following two classes.
\begin{itemize}
\item[i)] $\xi^\mu(w)$ is a  pull-back of a target space function 
$\xi^\mu (w)= \xi^\mu [X(w)]$.
It is related to the variation of the functional under diffeomorphism 
$X^\mu(z) \rightarrow X^\mu(z) +\xi^\mu[X(z)]$ as
$\delta_\xi F[X] = -\xi\act F[X]$.
\item [ii)] $\xi^\mu (w)$ is a function of $w$ but independent of $X(w)$ and its derivatives.
It corresponds to a change of the embedding, 
$X^\mu(z) \rightarrow X^\mu(z) +\xi^\mu(z)$, and is
used to derive the equation of motion.
\end{itemize}
By successive transformations, $\xi \triangleright (\eta \triangleright F)$, 
the vector fields $\xi$ and $\eta$ 
form a Lie algebra with the Lie bracket
\be
\label{Lie algebra}
[\xi, \eta]= \int \! d^2w \, \left(
\xi^\mu \frac{\delta \eta^\nu}{\delta X^{\mu}}
-\eta^\mu \frac{\delta \xi^\nu}{\delta X^{\mu}}
\right) (w) \frac{\delta}{\delta X^{\nu}(w)}~~.
\ee
Now we can define the universal 
enveloping algebra $\CH=U({\mathfrak X})$ of ${\mathfrak X}$ over $\complex$,
which has a natural cocommutative Hopf algebra structure
$(U({\mathfrak X}); \mu,\iota,\Delta,\epsilon,S)$. 
The defining maps given on elements $\xi, \eta \in {\mathfrak X}$ are
\begin{eqnarray}
&& \mu (\xi \otimes \eta )=\xi \cdot \eta~ , ~~~~ 
\iota(k)=k\cdot 1~, \nonumber \\ 
&&\Delta(1)=1\otimes 1~, ~~~\Delta(\xi )=\xi  \otimes 1+1 \otimes \xi~,  \nonumber \\ 
&&\epsilon(1)=1~,  ~~~\epsilon(\xi)=0~, \nonumber \\ 
&&S(1)=1, ~~~S(\xi )=-\xi~, 
\end{eqnarray}
where $k\in\complex$. 
As usual, these maps are uniquely extended to any such element of $U({\mathfrak X})$ by the 
algebra (anti-) homomorphism. 
In particular, the Poincar\'e-Lie algebra $\CP$ generated by 
\begin{eqnarray}
&&P^{\mu}=-i \int d^2z \,\eta^{\mu\lambda}
\frac{\delta}{\delta X^{\lambda}(z)},\nonumber \\ &&
L^{\mu\nu}=-i \int d^2z  X^{[\mu}(z) \eta^{\nu]\lambda} 
\frac{\delta}{\delta X^{\lambda}(z)}, 
\label{generators}
\end{eqnarray}
where $P^\mu$ are the generators of the translation and $L^{\mu\nu}$ are 
the Lorentz generators,
forms a Hopf subalgebra of $\CH=U({\mathfrak X})$, denoted as $U(\CP)$.
We denote the abelian Lie subalgebra consisting of $\xi$ 
(\ref{gene vector field}) in class ii) 
as ${\mathfrak C}$. 
Then $U({\mathfrak C})$ is also a Hopf subalgebra of $\CH$.

The algebra $\CA$ of functionals is now considered as a 
$\CH$-module algebra.
The action of an element $h \in \CH$ on $F \in \CA$ is denoted by
$h \triangleright F$, as above. 
The action on the product of two elements $F,G \in \CA$ is defined by 
\be
h \triangleright m (F\otimes G) = m\,\Delta(h) \triangleright (F\otimes G)~,
\label{covariance}
\ee 
which represents the covariance of the module algebra $\CA$ under 
diffeomorphisms or worldsheet variations.

\subsection{Quantization as a Hopf algebra twist}
\label{sec:twist quantization}

In Ref.\cite{AMW} we gave a simple quantization procedure in terms of a 
Hopf algebra twist,
in which the vacuum expectation value (VEV) coincides with the conventional 
path integral average.
A general theory of Hopf algebra twist is presented in Ref.\cite{2000fqgt.book.....M} 
and for details relating to our approach see also the Appendix in Ref.\cite{AMW}.

Suppose that there is a twist element
(counital 2-cocycle), $\CF \in \CH\otimes \CH$, which is
invertible, counital $({\rm id} \otimes \epsilon)\CF=1$ 
and satisfies the 2-cocycle condition
\bea
&&(\CF \otimes {\rm id})(\Delta \otimes {\rm id})\CF=({\rm id}\otimes\CF)
({\rm id} \otimes \Delta)\CF~.
\label{2-cocycle condition}
\eea
In this paper it is sufficient to assume that a twist element
is in the abelian Hopf subalgebra $\CF \in U({\mathfrak C})\otimes U({\mathfrak C})$ 
of the form
\begin{equation}
\CF=\exp\left(-{\int \!d^2z \!\!\int \!d^2w \,
G^{\mu\nu}(z,w)
\frac{\delta}{\delta X^{\mu}(z)}
\otimes\frac{\delta}{\delta X^{\nu}(w)}}\right),
\label{eq:F}
\end{equation}
which is specified by a Green function $G^{\mu\nu}(z,w)$ on the worldsheet.
It is easy to show that this $\CF$ in (\ref{eq:F}) satisfies all conditions 
for twist elements\cite{AMW}.

Given a twist element $\CF$, the twisted Hopf algebra $\CH_{\CF}$ can be
defined using the algebra and counit of the untwisted $\CH$, 
but with coproduct and antipode twisted
\be
\Delta_{\CF}(h)=\CF \Delta (h) \CF^{-1}, \quad 
S_{\CF}(h)=US(h)U^{-1}
\ee
for all $h \in \CH$, where $U=\mu ({\rm id}\otimes S)\CF$.
We regard this procedure of twisting as 
a quantization with respect to the twist element $\CF$.

Correspondingly, the consistency of the action, i.e. covariance with 
respect to the Hopf algebra action, 
requires that a $\CH$-module algebra $\CA$ is twisted 
to the $\CH_{\CF}$-module algebra $\CA_{\CF}$.
As a vector space, $\CA_{\CF}$ is identical to $\CA$,  
but is accompanied by the twisted product
\be
m_{\CF} (F\otimes G) =m\circ \CF^{-1}\act (F\otimes G)~.
\label{twistedProduct}
\ee 
This twisted product is associative owing to 
the cocycle condition (\ref{2-cocycle condition}).
We also denote it as $F*_{\CF} G$ in a more familiar notation, 
using the star product, as a result of a quantization.
See Ref.\cite{AMW} for remarks about the similarities and differences 
with deformation quantization.
Note that the twisted product remains commutative.
Since each element in $\CH_{\CF}$ as well as in $\CA_{\CF}$ 
is the same as the corresponding classical element, 
the variation of the local functional has 
the same representation, $h \,\act F[X]$, as the classical transformation.
However, the Hopf algebra action is not the same as the classical one 
when $I[X]$ is a product 
of several local functionals, since
\bea
 h\act m_{\CF} (F\otimes G)
&=& m\circ \Delta (h) \CF^{-1}\act (F\otimes G) \nonumber\\
&=& m_{\CF} \Delta_{\CF} (h) \act (F\otimes G)~.
\label{twisted covariance}
\eea
In this way the Hopf algebra and the module algebra are twisted in a consistent manner.
The resulting twisted action is considered as a transformation 
in the quantized theories.

We define the VEV for the twisted module algebra $\CA_{\CF}$ 
as the map $\tau: \CA_{\CF} \rightarrow \complex$.
For any element $I[X] \in \CA_{\CF}$ it is defined as the evaluation at $X=0$:
\be
\tau\left(\,I[X]\, \right):=I[X]|_{X=0}.
\label{def of true VEV}
\ee
If $I[X]$ is a product of two elements $F,G\in \CA_{\CF}$,
their correlation function is
\be
\sigma(z,w)=\tau(F[X(z)]*_{\CF}G[X(w)])
=\tau\circ m\circ \CF^{-1} \act (F\otimes G),
\ee 
Owing to the associativity,
the correlation function of $n$ local functionals is similarly given as
\be
\sigma(z_1,...,z_n)=\tau(F_1[X(z_1)]*_{\CF}F_2[X(z_2)]\cdots *_{\CF}F_n[X(z_n)]).
\ee 
The action of $h\in\CH_{\CF}$ inside the VEV $\tau (I[X])$ is
\be
\tau \left(h \act I[X]\right)
\label{Hopf action}
\ee
and this action appears in the various relations related with the symmetry transformation.

A different choice of the twist element $\CF$ gives a different 
quantization scheme. This completes the description of the procedure.

One advantage of this procedure is that both, the module algebra $\CA$ (observables) 
and the Hopf algebra $\CH$ (symmetry) are simultaneously quantized 
by a single twist element $\CF$.
This gives us a significantly simple understanding of the symmetry structure 
after the quantization,
as discussed in Ref.\cite{AMW} in great detail:
There, we argued that the symmetry 
of the theory in the conventional sense 
is characterized as a twist invariant Hopf subalgebra of $\CH_{\CF}$,
which consists of elements $h \in \CH$ such that $\Delta_{\CF}(h)=\Delta(h)$ 
and $S_{\CF}(h)=S(h)$.   
The other elements of $\CH_{\CF}$, i.e., generic 
diffeomorphisms, should be twisted
at the quantum level.
Since each choice of twist element chooses a particular background
(as we will see below when choosing an explicit $\CF$),
$\CH_{\CF}$ and $\CA_{\CF}$ are background dependent as well, but only through $\CF$
in the coproduct $\Delta_{\CF}$ or in the twisted product $m_{\CF}$.

The action of this twisted diffeomorphism can be 
regarded as a remnant of the classical diffeomorphism,
where the change of the background under diffeomorphisms is incorporated 
into the twisted diffeomorphism in such a way that 
the twist element (quantization scheme) is kept  invariant.

\subsection{Normal ordering and path integrals}
\label{sec:normal ordering}

Here we argue that the twist quantization described 
in the previous subsection is identical with
the path integral quantization, by identifying the 
corresponding VEVs.
As proved in Ref.\cite{AMW}, the twist element $\CF$ in (\ref{eq:F}) 
can be written as
\bea
&&\CF= \p \CN ^{-1}=(\CN ^{-1}\otimes \CN ^{-1})\Delta (\CN )~,\nonumber \\
&&\CF^{-1}=\p \CN = \Delta (\CN ^{-1}) (\CN  \otimes \CN )~,
\label{coboundaryrelation}
\eea
where $\CN \in \CH$ is defined by using the same Green's function as
 in $\CF$:
\be
\CN  = \exp\left\{-\frac{1}{2}\int \!d^2z \!\!\int \!d^2w \,
G^{\mu\nu}(z,w)
\frac{\delta}{\delta X^\mu(z)}\frac{\delta}{\delta X^\nu(w)}\right\} ~.
\label{normalorderingoperator}
\ee
This shows that the twist element $\CF$ is a coboundary 
and thus it is trivial in the Hopf algebra cohomology.
Consequently, there is an isomorphism between the Hopf algebras
 $\hat{\CH}$ and $\CH_{\CF}$ 
and the module algebras $\hat{\CA}$ and $\CA_{\CF}$, respectively, 
which can be summarized as:
\bea
\begin{array}{ccccc}
\CH & \xrightarrow{\mbox{twist}} & \CH_{\CF} & \xrightarrow{\sim} & 
\hat{\CH} \\
\triangledown & & \triangledown & & \triangledown  \\
\CA & \xrightarrow{\mbox{twist}} & \CA_{\CF} & \xrightarrow{\sim} 
& \hat{\CA}  
\end{array}
\label{diagram}
\eea
In the diagram, the left row is the classical pair $(\CH,\CA)$, 
and the middle and the right rows are the quantum counterparts.
Here the map $\CH_{\CF} \xrightarrow{\sim} \hat{\CH}$ is given by an 
inner automorphism $h \mapsto \CN  h {\CN }^{-1} \equiv \tilde{h}$, and 
the map $\CA_{\CF} \xrightarrow{\sim} \hat{\CA}$ is given by 
$F \mapsto \CN  \act F \equiv :\!F\!:$ (see below).
We call $\hat{\CH}$ ($\hat{\CA}$) the normal 
ordered Hopf algebra (module algebra), respectively.

The normal ordered module algebra $\hat{\CA}$ is the one 
which appears inside the VEV in the path integral.
It consists of elements in normal ordered form
\be
\CN  \act I[X] \equiv :\!I[X]\!:
\label{normal ordered functional}
\ee 
for any functional $I[X] \in \CA_{\CF}$. 
These elements correspond to vertex operators in the path integral.
Note that the action of $\CN$ (\ref{normalorderingoperator}) corresponds to 
subtractions of divergences at coincident points caused by self-contractions 
in the path integral.
As an algebra, $\hat{\CA}$ has the same multiplication 
$m:\hat{\CA}\otimes\hat{\CA}\rightarrow \hat{\CA}$
as the classical functional $\CA$.
This is seen by the map of the product in $\CA_{\CF}$ as
\be
 \CN  \act m \circ \CF^{-1} \act (F\otimes G) 
= m \circ (\CN  \otimes \CN )\act (F\otimes G)~,
\label{product iso}
\ee
which is a direct consequence of the coboundary relation (\ref{coboundaryrelation}).
An equivalent but more familiar expression, 
$:\! (F *_{\CF} G) \!: = :\!F\!:\, :\!G\!:$,
is simply the time ordered product of two vertex operators.

The normal ordered Hopf algebra $\hat{\CH}$ is defined with the same 
algebraic operations as the classical Hopf algebra $\CH$, but 
with elements dressed with a normal ordering as
\be
\CN  h {\CN }^{-1} \equiv \tilde{h}
\label{normal ordered h}
\ee 
for any $h \in \CH_{\CF}$. 
The isomorphism map relates the twisted Hopf algebra action $\CH_{\CF}$ on 
$\CA_{\CF}$ to the corresponding action of $\hat{\CH}$ on $\hat{\CA}$.
For example,
\bea
h\act F ~\xrightarrow{\sim}~ 
\CN  \act (h \act F) =\,
\tilde{h}~\act :\!\!F\!: \label{action on :F:2}\\
h\act (F*_{\CF} G) \xrightarrow{\sim}\tilde{h} \act (:\!F\!::\!G\!:).
\label{action on :F:}
\eea

As argued in Ref.\cite{AMW}, some elements in $\hat{\CA}$
contain the formally divergent series which we have to deal 
with in the functional language 
(this is the reason why we distinguish $\CA$ and $\hat{\CA}$),
and thus eqs.\,(\ref{action on :F:2}) and (\ref{action on :F:}) have only a meaning inside the VEV. 
The VEV (\ref{def of true VEV}) for $\CA_{\CF}$ implies the 
definition of the VEV 
for $\hat{\CA}$ to be a map, $\tau \circ \CN ^{-1}: \hat{\CA}\rightarrow \complex$, 
and it turns out that it coincides with the VEV $\langle \cdots \rangle$ 
in the path integral. 
For instance, the correlation of two local functionals is
\be
\tau \circ \CN ^{-1} \act (:\!F[X]\!:\!(z) :\!G[X]\!:\!(w))
=\langle\, :\!F[X]\!:\!(z) :\!G[X]\!:\!(w) \,\rangle 
\label{VEVpathintegral}
\ee
Here, the action $\CN ^{-1}$ gives the Wick contraction with respect to the Green function.
The equality (\ref{VEVpathintegral}) can be easily 
verified using the standard path integral argument (see for example Ref.\cite{Polchinski}).
The VEV $\langle \cdots \rangle$ in the path integral means 
\begin{eqnarray}
\langle \CO\rangle :=
\frac{\int \CD X \CO e^{-S}}{\int \CD X e^{-S} }~.
\label{VEV}
\end{eqnarray} 
Here the action functional $S[X]$ is related to the choice of the 
Green function.
It is quadratic $S[X]=\frac{1}{2}\int d^2z X^\mu D_{\mu\nu} X^\nu$
with a fixed second order derivative $D_{\mu\nu}$ such that 
the Green function in (\ref{normalorderingoperator}) is a solution of 
the equation 
$D_{\mu\rho}G^{\rho\nu}(z,w) =\delta_\mu^\nu \delta^{(2)}(z-w)$. 
Note that in (\ref{VEV}), each local insertion is understood as being 
regulated 
by the (conformal) normal ordering $\CN$.
This must correspond to a (operator) normal ordering in the operator formulation
\footnote{It is also possible to define a path integral with 
regularization 
prescriptions other than the conformal normal ordering used here.
In that case, the corresponding operator ordering is also changed.
We do not exclude this possibility but claim that there is a natural choice 
of the normal ordering from the viewpoint of classical functionals,
in the sense that the functionals themselves are 
not modified under the quantization.}.

The two descriptions in terms of a twisted pair $(\CH_{\CF}, \CA_{\CF})$ 
and 
of a normal ordered pair $(\hat{\CH},\hat{\CA})$ (thus path integral) are 
equivalent. 
However, note that the background dependences in the 
two formulations are different.
In the case of the normal ordered pair $(\hat{\CH},\hat{\CA})$, 
both, an element $:\!F\!:\in \hat{\CA}$ and the VEV $\tau \circ \CN ^{-1}$, 
contain $\CN$, which depends on the background.
This corresponds, in the operator formulation, to the property that 
a mode expansion of the string variable $X^\mu(z)$ as well as 
the oscillator vacuum are background dependent.
Therefore, the description of the quantization 
that makes $\hat{\CA}$ well-defined is 
only applicable to that particular background and we need a 
different mode expansion for a different background.

\section{Twisted Hopf algebra in $B$ field background}
\label{sec:B}

In this section, we study the Hopf algebra structure 
for the case of a non-zero $B$-field background, following the above 
quantization procedure based on the Hopf algebra twist. 
The relation between the quantization 
and twist as explained in section \ref{sec:A} 
leads to another quantization scheme for 
the case with $B$-field background, where, 
we need, in particular, a deformed normal ordering.

\subsection{String worldsheet theory in $B$ field background}
\label{sec:worldsheet}

Consider the bosonic closed strings as well as open strings and 
take a space-filling D-brane for simplicity.
We start with the $\sigma$ model of the bosonic string 
with flat $d$-dimensional Minkowski space as the target.
The action in the conformal gauge is 
\be
S_0 [X]=\frac{1}{2\pi \alpha'}\int_{\Sigma} d^2z \eta_{\mu\nu}\p X^\mu \bar\p X^\nu ,
\label{action}
\ee
where the world sheet $\Sigma$ 
can be any Riemann surface with boundaries, and 
typically, we take it as the complex plane (upper half plane) 
for a closed string (open string).
$z^a=(z, \bar{z})$ are the complex coordinates on the world sheet. 
The flat metric in the target space $\real^d$ is represented
by $\eta^{\mu\nu}$.

The constant $B$ field is turned on by adding a term
\be
S_B=\frac{1}{2\pi \alpha'}\int_{\Sigma} d^2z B_{\mu\nu}\p X^\mu \bar\p X^\nu 
\label{S_B}
\ee
to the free action (\ref{action}). This 
does not affect the equation of motion for $X^\mu(z)$, but 
it changes the boundary condition from Neumann boundary condition to the mixed one.
Therefore, we consider the worldsheet with boundaries in this section.
To be more explicit, the action of the theory is given by
\bea
S_1 &=& S_0+S_B \nonumber\\
&=& \frac{1}{2\pi\alpha'} \int_{\Sigma} d^2z (\eta_{\mu\nu}+ B_{\mu\nu})
\p X^\mu \bar\p X^\nu \nonumber\\
&=& \frac{-1}{2\pi\alpha'} \int_{\Sigma} d^2z \eta_{\mu\nu} X^\mu \p \bar\p X^\nu 
+ \frac{1}{2\pi\alpha'} \int_{\p\Sigma} dt
X^{\mu} (\eta_{\mu\nu} \p_n X^\nu + B_{\mu\nu} \p_t X^\nu ),
\label{S_1}
\eea
where integration by parts is used.
$\p\Sigma$ is in general a sum of boundary components of the 
worldsheet $\Sigma$, and $t$ is its boundary coordinate. 
$\p_n$ is the normal derivative to $\p\Sigma$, 
and $\p_t$ is the tangential derivative along $t$, respectively.

With the constant $B$-field background, the propagator
can be calculated by solving the equation of motion
on the worldsheet with the mixed boundary condition and 
we denote it by the subscript $1$ according to $S_1$:
\be
\langle X^{\mu}(z) X^{\nu}(w) \rangle_1 
= G_1^{\mu\nu}(z,w).
\label{G_1}
\ee
This propagator satisfies 
\be
-\frac{1}{\pi\alpha'} \p\bar{\p} G_1^{\mu\nu} (z,w) = \eta^{\mu\nu}\delta^{(2)}(z-w),\quad
(\eta_{\mu\nu} \p_n  + B_{\mu\nu} \p_t )G_1^{\mu\nu} (t,w)=0.
\label{Green relation with B}
\ee
In the simplest case, $\Sigma$ is the upper half plane of the 
complex plane where $G_1^{\mu\nu}$ is given by 
Refs.\cite{Fradkin:1985qd,1987NuPhB.288..525C,1987NuPhB.280..599A}
\begin{align}
G_1^{\mu\nu}(z,w) = -{\alpha'} & \Big[ 
\eta^{\mu\nu}\ln |z-w| -\eta^{\mu\nu}\ln |z-\bar w|  \nonumber\\
&  +G^{\mu\nu}\ln |z-{\bar w}|^2 +  {\Theta^{\mu\nu}} \ln 
\frac{z-{\bar w}}{{\bar z}-w} \Big] ,  
\label{UHPpropagator}
\end{align}
where the open string metric $G^{\mu\nu}$ and the antisymmetric tensor $\Theta^{\mu\nu}$ are 
defined as 
\be
G^{\mu\nu}
=\left(\frac{1}{\eta+B} \eta \frac{1}{\eta-B} \right)^{\mu\nu}
\and \Theta^{\mu\nu}= \frac{\theta^{\mu\nu}}{2\pi \alpha'} 
= -\left(\frac{1}{\eta+B} B \frac{1}{\eta-B} \right)^{\mu\nu}.
\ee
If we set $B=0$, (\ref{G_1}) reduces to the propagator $G_0^{\mu\nu}(z,w)$ 
in Ref.\cite{AMW} which satisfies the
Neumann boundary condition $\p_n G_0^{\mu\nu} (t,w)=0$.

\subsection{Twist quantization with $B$ field}
\label{sec:H_1}

Let $\CH=U({\mathfrak X})$ be the Hopf algebra of functional variations 
and let $\CA$ be its module algebra given by the classical 
functionals of the embedding $X$.
In the $B$ field background, the total action of the theory is 
$S_1=S_0+S_B$, such that the propagator is $G_1^{\mu\nu}(z,w)$.
Then, in our approach, the quantization of the system is defined 
by the twist element
\be
\CF_1:=\exp\left\{-\int \! d^2z \!\! \int \! d^2w \,
G_1^{\mu\nu}(z,w)
\frac{\delta}{\delta X^\mu(z)}
\otimes \frac{\delta}{\delta X^\nu(w)}\right\}~~.
\label{F_1}
\ee
The whole structure concerning the quantization is exactly the same as in 
the previous section.
Since $\CF_1$ satisfies the unital 2-cocycle condition,
we obtain the twisted Hopf algebra $\CH_{\CF_1}$ and its module algebra 
$\CA_{\CF_1}$ as a quantization of $\CH$ and $\CA$. 
As discussed in section \ref{sec:normal ordering}, 
there are (formal) isomorphisms $\CH_{\CF_1}\simeq \CH$ and 
$\CA_{\CF_1}\simeq \CA$ relating the twisted Hopf (module) algebras and the normal ordered
Hopf (module) algebras.
In order to distinguish them from the classical ones, 
we denote them by $\hat{\CH}_1$ and $\hat{\CA}_1$, respectively.
They consist of elements of the form 
$\tilde{h}=\CN_1 h\CN_1^{-1} \in \hat{\CH}_1$ and 
$\CN_1 \act F \in \hat{\CA}_1$,
respectively, as in the general construction in section \ref{sec:normal ordering}.
Here, the normal ordering element is
\begin{eqnarray}
\CN_1=\exp\left\{-\half\int d^2z d^2w ~G_1^{\mu\nu}(z,w)
\frac{\delta}{\delta X^{\mu}(z)} \frac{\delta}{\delta X^{\nu}(w)}\right\}.
\label{N_1}
\end{eqnarray}  
We denote this normal ordering 
as $\,^{\circ}_{\circ} \cdots \,^{\circ}_{\circ} $ and the star product 
as $F*_{\CF_1}G = m\CF_1^{-1} \act (F\otimes G)$, then their relation
analogous to \eqref{product iso} is written as
\be
\,^{\circ}_{\circ} F*_{\CF_1}G \,^{\circ}_{\circ} 
=\,^{\circ}_{\circ} F \,^{\circ}_{\circ} \,^{\circ}_{\circ} G \,^{\circ}_{\circ} 
\ee
The VEV for a normal ordered functional 
$\,^{\circ}_{\circ}I[X]\,^{\circ}_{\circ} \in \hat{\CA}_1$ is
\be
\langle \,\,^{\circ}_{\circ}I[X]\,^{\circ}_{\circ}\,\rangle_1
=\tau\left(\CN_1^{-1}
\triangleright \,^{\circ}_{\circ}I[X]\,^{\circ}_{\circ} \right)
=\tau \left(I[X]\right),
\ee
which coincides with the path integral average with respect to the action $S_1$
\begin{eqnarray}
\langle \CO\rangle _1=
\frac{\int \CD X \CO e^{-S_1}}{\int \CD X e^{-S_1} }.
\label{path integral}
\end{eqnarray}

Note that here, for the case 
with the background $(\eta_{\mu\nu},B_{\mu\nu})$, we denote the twist 
element as $\CF_1$, 
and the related structures such as $\CH_{\CF_1}$, $\CA_{\CF_1}$, 
$\CN_1$ and $*_{\CF_1}$ with a suffix $1$.
On the other hand, the twist quantization in the background 
$(\eta_{\mu\nu},B_{\mu\nu}=0)$ as 
treated in Ref.\cite{AMW}
is determined by the twist element $\CF_0$ with the propagator $G_0^{\mu\nu}(z,w)$.
Following this notation we denote the resulting structures such 
by $\CH_{\CF_0}$, $\CA_{\CF_0}$, 
$\CN_0$ and $*_{\CF_0}$ with a suffix $0$.
The normal ordering in this case is denoted as 
$\CN_0 \,\act F =:\!F\!:$ and the normal ordered Hopf 
and module algebras are denoted by $\hat{\CH}$ and $\hat{\CA}$, 
respectively.\footnote{
In principle, all the formulas relating the quantization and
normal ordering in Ref.\cite{AMW} is valid for both cases with suffix $0$ or $1$.}.

Comparing with the $B=0$ case, the change of the background to a 
non-zero value 
affects directly the normal ordered 
module algebra $\hat{\CA}_1$, which is in contrast to the 
twisted counterpart $\CA_{\CF_1}$, because an element of $\hat{\CA}_1$
has a different power series 
expansion, i.e., $\CN_1 \act F$, rather than $\CN_0 \act F$ for the case without 
$B$-field background in Ref.\cite{AMW}.
This modification of the normal ordering
 corresponds to the modification of the operator description,
where a new mode expansion and a new Fock vacuum are required by the change of background.
In fact, the normal ordering defined by \eqref{N_1} coincides with that 
proposed in 
Refs.\cite{Braga:2004wr,Braga:2006gi,Chakraborty:2006yj,Gangopadhyay:2007ne}.
There, this normal ordering was introduced such that a normal ordered operator like 
$\,^{\circ}_{\circ}\hat{X}^\mu (z)\hat{X}^\nu (w)\,^{\circ}_{\circ}$
satisfies not only the equation 
of motion, but also 
the mixed boundary condition as an operator relation. 
The authors also observed the noncommutativity of position 
operators by an appropriate mode expansion.
Our results here give a foundation of this proposal.

\paragraph{Twisted symmetry with $B$ field.}
The discussion of the symmetry in the case with $B$-field background 
goes analogously to the case without $B$ field studied in Ref.\cite{AMW}. 
Using the above definition of the VEV, 
the formula of the twisted Hopf algebra $\CH_{\CF_1}$ action on a functional inside the VEV
can be written as
\bea
\langle\, \tilde{h} ~\act \,^{\circ}_{\circ}F[X]\,^{\circ}_{\circ} \,\rangle_1
&=&
\tau \left( h \act F[X] \right)~.
\label{H_1 action for F}
\eea
where $\tilde{h}=\CN_1 h\CN_1^{-1} \in \hat{\CH}_1$.
We can see that the key relation 
\be
\xi \act I[X] = -m \circ (\xi \otimes 1) F_1 
\act \left(S_1 \otimes I[X] \right),
\label{xI=xFSI with B}
\ee
for deriving the Ward-like identities holds as in the case 
without $B$ field (See Ref.\cite{AMW}). 
Here, $F_1$ is defined by $\CF_1=e^{F_1}$.
The proof of (\ref{xI=xFSI with B}) is as follows:
 The action of $F_1$ on $S_1 \otimes I[X]$ is 
written using (\ref{S_1}) as
\begin{align}
F_1 \act &\left(S_1 \otimes I[X] \right)
= -\frac{1}{\pi\alpha'}
\int \!\! d^2 z \!\!\int \!\! d^2 w\, G_1^{\mu\nu} (z,w) 
\left( \p\bar{\p} X_\mu (z) \otimes 
\frac{\delta I[X] }{\delta X^\nu (w)} \right) \\
& + \frac{1}{2\pi\alpha'}\int_{\p\Sigma}\!\! dt \!\!\int \!\! d^2 w\, G_1^{\mu\nu} (t,w) 
\left(
(\eta_{\mu\nu} \p_n X^\nu + B_{\mu\nu} \p_t X^\nu )(t)
\otimes 
\frac{\delta I[X]}{\delta X^\nu (w)} \right). \nonumber
\end{align}
Integrating by parts, and using the defining relations of the Green function
(\ref{Green relation with B}),
we confirm (\ref{xI=xFSI with B}).
In particular, we obtain the Ward-like identity
analogous to eq.(43) in Ref.\cite{AMW} as
\be
0= \langle ~\tilde{\xi}~ \act \,^{\circ}_{\circ} I[X]\,^{\circ}_{\circ} \rangle_1 
- \langle \,^{\circ}_{\circ}(\xi \act S_1)\,^{\circ}_{\circ} 
\ \,^{\circ}_{\circ}I[X]\,^{\circ}_{\circ} \rangle_1
- m \circ \left[\xi \otimes 1, \CF_1^{-1} \right]
\act \left(S_1 \otimes I[X] \right) ~.
\label{general identity for Hopf2 with B}
\ee
Since a infinitesimal Lorentz transformation in a generic $B$-field background 
does not preserve the action: $\xi \act S_1 \ne 0$, 
nor does the commutator vanishes: $[\xi \otimes 1, \CF_1^{-1}] \ne 0$,
we have only broken-type Ward identities.
This feature is equivalent to the statement that the Poincar\'e symmetry as well as 
the full space-time diffeomorphism should be twisted at the quantum 
level.\footnote{Of course, when some of component of $B_{\mu\nu}$ happens to be zero,
an unbroken symmetry remains.}
It will become clearer in the next section, 
how the original Poincar\'e symmetry is deformed by the $B$ field.

\section{Decompositions of the twist and physical interpretations}
\label{sec:relation}

We have shown that application of the twist quantization to the case with nontrivial $B$-field background gives a twisted Hopf algebra including the diffeomorphism of the target space. Thereby, the quantization and the deformation by the $B$-field are treated in a unified way, which was one of our aims. On the other hand, this unified approach makes the geometrical implication of the deformation by the $B$-field background 
less transparent. 

For the open string case, it is well known that the effective theory 
with the $B$-field background can be formulated as a field theory on the 
noncommutative space. The effect of the $B$-field is reflected only on the noncommutativity and the corresponding geometrical picture is noncommutative space. On the other hand, for the 
closed string, such a geometrical interpretation of the $B$-field is 
missing. It is an interesting question whether we can have 
an appropriate geometrical structure for the closed string case. 

In the following, we develop a method to decompose 
the twist into two successive twists, and discuss its interpretation.
We propose two different decompositions 
of the twist and discuss the feature of the 
$B$-field effect as a deformation of the functional algebra and as 
a twist of the symmetry, respectively. 

The first is a natural decomposition from the viewpoint of 
 the twist quantization.
There we can see the relation 
between the standard quantization and the deformation 
caused by a $B$-field background.
With this decomposition, we obtain a new twisted Poincar\'e symmetry, 
which is different from the one considered in Refs.\cite{Chaichian:2004za,Koch:2004ud, Aschieri:2005yw}. The twisted Poincar\'e symmetry
obtained here should be considered as a symmetry structure underlying 
the string theory. 

However, the first decomposition becomes singular when we take the 
Seiberg-Witten limit\cite{Seiberg:1999vs} to obtain the noncommutative field theory picture 
of the effective theory.
The second decomposition is meaningful 
for the open string case. By restricting the functional space to the
one corresponding to the open string vertex operators,
 we can derive the relation to the 
twisted Poincar\'e symmetry in the field theory on the noncommutative 
space considered in Refs.\cite{Chaichian:2004za,Koch:2004ud, Aschieri:2005yw}.

\subsection{$B$ field as a deformation}
\label{sec:2step twist}

Using the twist quantization, the effect of the $B$ field background
is unified with the quantization twist into a single twist element $\CF_1$.  
Therefore, to see the effect of the $B$ field background separately, 
we have to extract it from $\CF_1$.
It is also more natural to compare directly the twisted Hopf 
algebras $\CH_{\CF_1}$ and $\CH_{\CF_0}$, 
without referring to the classical Hopf algebra $\CH$.

\paragraph{Successive twists.}
For this purpose, we consider a decomposition
of the propagator (\ref{G_1}) into the following two parts:
\be
\langle X^{\mu}(z) X^{\nu}(w) \rangle_1 
= G_1^{\mu\nu}(z,w) = G_0^{\mu\nu}(z,w) + G_B^{\mu\nu}(z,w)~,
\label{G_1=G_0+G_B}
\ee
where $G_0^{\mu\nu}$ is the propagator in $B=0$.
Then, the propagators are given, on the upper half plane, for 
example, by
\bea
&& G_0^{\mu\nu}(z,w):=-{\alpha'}\eta^{\mu\nu}(\ln |z-w|+\ln |z-\bar w|) \nonumber\\
&& G_B^{\mu\nu}(z,w):=-{\alpha'} \Big[ 
(G-\eta)^{\mu\nu} \ln |z-{\bar w}|^2 +  {\Theta^{\mu\nu}} \ln 
\frac{z-{\bar w}}{{\bar z}-w}\Big]~.
\label{propagatorwithB}
\eea

Accordingly, the twist element $\CF_1$ (\ref{F_1}) is also divided into 
\be
\CF_1=\CF_B \CF_0~,
\label{F_1=F_BF_0}
\ee
where $\CF_B$ is the pure effect of the $B$ field defined by 
\be
\CF_B:=\exp\left\{-\int \! d^2z \!\! \int \! d^2w \,
G_B^{\mu\nu}(z,w)
\frac{\delta}{\delta X^\mu(z)}
\otimes \frac{\delta}{\delta X^\nu(w)}\right\}~~.
\label{F_B}
\ee
The decomposition (\ref{F_1=F_BF_0}) defines two successive twists 
\be
\CH \xrightarrow{\text{twist by $\CF_0$}} \CH_{\CF_0} \xrightarrow{\text{twist by $\CF_B$}}
 (\CH_{\CF_0})_{\CF_B}=\CH_{\CF_1}~~.
\label{2step twist}
\ee
We give a remark for the last equality.
In general, for a given a twisted Hopf algebra $\CH_{\CF_0}$ with a 
coproduct $\Delta_{\CF_0}$, 
a further twisting by $\CF_B$ is possible if $\CF_B$ is a 2-cocycle in $\CH_{\CF_0}$ 
\be
(\CF_B \otimes {\rm id})(\Delta_{\CF_0} \otimes {\rm id})\CF_B =({\rm id}\otimes\CF_B)
({\rm id} \otimes \Delta_{\CF_0})\CF_B~.
\label{2nd cocycle condition}
\ee
Then, it is equivalent to the 1-step twist $\CH_{\CF_1}$, where $\CF_1=\CF_B \CF_0$. 
This property holds because the l.h.s. of the 2-cocycle condition 
(\ref{2-cocycle condition}) for $\CF_1$ 
can be written using (\ref{2-cocycle condition}) for $\CF_0$ and 
(\ref{2nd cocycle condition}) as
\be
(\CF_1 \otimes {\rm id})(\Delta \otimes {\rm id})\CF_1 
=\left[(\CF_B \otimes {\rm id})(\Delta_{\CF_0} \otimes {\rm id})\CF_B\right]
 \,\left[(\CF_0 \otimes {\rm id})(\Delta \otimes {\rm id})\CF_0 \right],
\ee
and similarly for the r.h.s..
Conversely, if two twist elements $\CF_0$ and $\CF_1$ in $\CH$ have a relation 
$\CF_1=\CF_B \CF_0$,
then $\CF_B$ is a twist element in $\CH_{\CF_0}$ satisfying (\ref{2nd cocycle condition}).
Note that in our case, the decomposition of the Green function is directly related to the 
decomposition of the twist element, because these are all abelian twists and
 (\ref{2nd cocycle condition}) holds.
Therefore, the effect of the $B$-field is completely characterized 
as a (second) twist $\CF_B$ of the Hopf algebra $\CH_{\CF_0}$ 
and the module algebra $\CA_{\CF_0}$.

Using the isomorphism (\ref{diagram}), 
$\CH_{\CF_0}$ can be replaced with the normal ordered 
$\hat{\CH}$ with the standard normal ordering $\CN_0$.
Of course, there is also an isomorphism $\CH_{\CF_1}\simeq \hat{\CH}_1$ 
as described in section \ref{sec:H_1}.
In this case,  the second twist by $\CF_B$ in (\ref{2step twist}) 
can also be regarded as a twist 
relating the two normal ordered Hopf algebras 
$\hat{\CH}\rightarrow \hat{\CH}_{\CF_B}=\hat{\CH}_1$.
Correspondingly, we have a map of normal ordered module algebras 
$\hat{\CA}\rightarrow \hat{\CA}_{\CF_B}=\hat{\CA}_1$ for the second twist.
This enables us to study the effect of the $B$-field in terms of vertex operators
as we will see in the following.

\paragraph{$B$ field as a deformation of OPE.}
First, we compare the two normal ordered functionals $:\!F\!: \in \hat{\CA}$
and $\,^{\circ}_{\circ} F \,^{\circ}_{\circ} \in \hat{\CA}_1$.
They correspond to operators in two different quantization schemes.
According to (\ref{F_1=F_BF_0}), the normal orderings are related as
$\CN_1=\CN_B \CN_0$, which are defined in the same manner as (\ref{coboundaryrelation}).
Then, we have a relation 
$\,^{\circ}_{\circ} F \,^{\circ}_{\circ} = \CN_B \,\act :\!F\!:$ between them.
The relation is such that the VEVs, each with respect to the corresponding 
quantization scheme, give the same result
\be
\langle\,^{\circ}_{\circ} F \,^{\circ}_{\circ} \rangle_1
= \tau \left( \CN_1^{-1} \act \,^{\circ}_{\circ} F \,^{\circ}_{\circ} \right)
= \tau \left( \CN_0^{-1} \act :\! F \!:\right)
= \langle\, :\!F\!: \,\rangle_0.
\ee
At first sight, it may be curious that the two different theories give the same VEV.
{However, this is a requirement}, since the definition of the VEV 
corresponds to the {\it normalized} path integral average 
(\ref{path integral}).
The difference of two quantization schemes lies in the Wick contraction 
of several local functionals.

Next, the twisted product of two local functional $F,G \in \CA_{\CF_1}$ is rewritten 
by using (\ref{F_1=F_BF_0}) in terms of 
the elements in $\hat{\CA}_{\CF_B}$ as
\bea
F*_{\CF_1} G 
&=& m \circ \CF_1^{-1} \act (F\otimes G) \nonumber \\
&=& m \circ \CF_B^{-1} \Delta(\CN_0^{-1})(\CN_0 \otimes \CN_0) \act (F\otimes G) \nonumber \\
&=& \CN_0^{-1} \act m \circ \CF_B^{-1} (\CN_0 \act F \otimes \CN_0 \act G) \nonumber \\
&=& \CN_0^{-1} \act \left(:\!F\!: *_{\CF_B} :\!G\!: \right),
\label{twist-normal relation}
\eea
where the coboundary relation (\ref{coboundaryrelation}) is used for $\CF_0^{-1}$.
Note that this simply means the equivalence of the two star products 
$:\!F*_{\CF_1} G\!:=:\!F\!: *_{\CF_B} :\!G\!:$.
By applying $\tau$ on both sides, this leads to the equation of the VEV 
\bea
\langle\,^{\circ}_{\circ} F \,^{\circ}_{\circ} \,^{\circ}_{\circ} G 
\,^{\circ}_{\circ} \,\rangle_1
&=&\tau (F*_{\CF_1} G) \nonumber \\
&=&\tau \circ \CN_0^{-1} \act \left(:\!F\!: *_{\CF_B} :\!G\!: \right) \nonumber \\
&=&\langle\, :\!F\!:*_{\CF_B} :\!G\!: \,\rangle_0.
\eea
On the l.h.s., the definition of the VEV as well as the normal 
ordering are
with respect to the new quantization $\hat{\CA}_1$ including the $B$ field.
The r.h.s. is written in terms of the standard quantization scheme 
$\hat{\CA}$, 
except that the OPE is twisted by $\CF_B$ (thus $\hat{\CA}_{\CF_B}$).

Now we relate the twisted Hopf algebra action to the normal ordered Hopf algebra action.
As above, the action of $h \in \CH_{\CF_1}$ on any functional 
$I \in \CA_{\CF_1}$ is related to 
the action of $\tilde{h}=\CN_0 h \CN_0^{-1} \in \hat{\CH}$ on $:\!I\!: \in \hat{\CA}$
as $h\act\, I = \CN_0^{-1} \tilde{h} \, \act :\!I\!:$, or equivalently, 
$\,^{\circ}_{\circ} h\act\, I \,^{\circ}_{\circ} 
=\CN_B \tilde{h} \,\act :\!I\!:$. 
Using (\ref{twist-normal relation}), 
we get for the product of two local functional $I=F*_{\CF_1} G$, 
\bea
\langle\,^{\circ}_{\circ} h \act \left(F *_{\CF_1} G \right)\,^{\circ}_{\circ} \,\rangle_1
&=&\tau \circ 
\CN_0^{-1} \tilde{h}\, \act \left(:\!F\!: *_{\CF_B} :\!G\!: \right)
 \nonumber \\
&=&\langle\, \tilde{h}\, \act \left(:\!F\!: *_{\CF_B} :\!G\!: \right) \,\rangle_0,
\eea
where the action of $\tilde{h}$ in the last line is written as 
\be
\tilde{h}\, \act \left(:\!F\!: *_{\CF_B} :\!G\!: \right) 
= m \circ \CF_B^{-1} \Delta_{\CF_B} (\tilde{h}) \act \left(:\!F\!: \otimes :\!G\!: \right).
\ee
This formula is a direct consequence of the second twist 
$\hat{\CH} \rightarrow \hat{\CH}_{\CF_B}$,
and shows that the transformation on the operator algebra deformed by the $B$ field should 
also be twisted.

In summary, at the level of vertex operators,
the quantization scheme with $B$ field (twist by $\CF_1$) can also be considered in 
the standard quantization scheme without $B$ field 
(by $\CF_0$), but with the operator product and the Hopf algebra 
action being twisted by the twist element $\CF_B$.

\paragraph{Twisted Poincar\'e symmetry.}
Here we discuss about the fate of the Poincar\'e symmetry $U(\CP)$.
As already observed in section \ref{sec:H_1},
the universal envelope of the Poincar\'e-Lie algebra $U(\CP)$ has 
to be twisted in a generic $B$-field background.
Note that this is true even for closed string vertex operators 
in the presence of the boundary $\p \Sigma\ne 0$.\footnote{Of 
course if $\p \Sigma= 0$, then $G_B$ can be set to $0$.}
By extracting the effect of the $B$-field as the second twist $\CF_B$, 
the structure of this twisted Poincar\'e symmetry becomes transparent as follows.

For $B=0$ case, $U(\CP)$ is a twist invariant Hopf subalgebra of $\CH$
under the twisting by $\CF_0$, as argued in Ref.\cite{AMW}.
Equivalently, $U(\CP)$ is a Hopf subalgebra of $\hat{\CH}$ 
with $\tilde{P}_{\mu}=P_{\mu}$ and $\tilde{L}_{\mu\nu}=L_{\mu\nu}$ 
as elements in $\hat{\CH}$. 
Thus, $U(\CP)$ remains a symmetry at the quantum level, and each (normal ordered) 
vertex operator in $\hat{\CA}$ is in a representation of a Poincar\'e-Lie algebra.
We emphasize that this fact guarantees 
the spacetime meaning of a vertex operator.
For example, a graviton vertex operator 
$V=:\!\p X^\mu \bar{\p}X^\nu e^{ikX}\!:$ transforms as spin $2$ representation 
under $U(\CP)$ corresponding to the spacetime graviton field $h_{\mu\nu}(x)$.

However, in a $B$-field background,
the second twist $\hat{\CH} \rightarrow \hat{\CH}_{\CF_B}$ 
modifies the coproduct and antipode of the 
Lorentz generator in 
$U(\CP)$ as
\begin{align}
&\Delta_{\CF_B} (L_{\mu\nu}) = L_{\mu\nu}\otimes 1+1\otimes L_{\mu\nu}
-2\int d^2 z d^2 w \, \eta_{[\mu\alpha} G_B^{\alpha\beta }(z,w)
\frac{\delta}{\delta X^\beta (z)}\otimes \frac{\delta}{\delta X^{\nu]} (z)} 
\label{twisted coproduct for L}\\
&S_{\CF_B}(L_{\mu\nu}) =S(L^{\mu\nu})+
2\int\!\!d^2ud^2z\,G_B^{\rho[\mu}(u,z)\frac{\delta}{\delta X^{\rho}(u)}
\frac{\delta}{\delta X_{\nu]}(z)} .
\label{twisted antipode for L}
\end{align}
Apparently, this $L_{\mu\nu}$ is not primitive and this means 
that the $U(\CP)$ is twisted.
We give several remarks about this twisted Poincar\'e symmetry read
from (\ref{twisted coproduct for L}), (\ref{twisted antipode for L}) 
and (\ref{F_B}):
\begin{enumerate}
\item
The twisting of $U(\CP)$ is only due to the second twist $\CF_B$.
This guarantees that a single local vertex operator $:\!F\!:$ 
in $\hat{\CA}_{\CF_B}$ is still Poincar\'e covariant after the twist
(in the same representation under $U(\CP) \subset \hat{\CH}$), 
and thus the spacetime 
meaning of a vertex operator 
(e.g., a graviton as spin $2$ representation) is unchanged.
\item 
The twisting is not closed within $U(\CP)$, 
but the additional terms in (\ref{twisted coproduct for L}) are 
in $U({\mathfrak C})\otimes U({\mathfrak C})$, 
and similarly for (\ref{twisted antipode for L}).
This shows that the twisting does not mix the Lorentz generator with other 
diffeomorphisms even for products of vertex operators.
Therefore, from the spacetime point of view, 
a Poincar\'e transformation on products of fields 
is still a global transformation even after the twist.
\item The twisted product $*_{\CF_B}$ of two vertex operators
reduces to the product of corresponding spacetime fields, 
but the latter depends on the representations (spins) of $U(\CP)$,
because $\CF_B$ in (\ref{F_B}) depends on the Green function $G_B$.
For instance, the twisted product of two gravitons is different from
that of two tachyons due to this fact.
Similarly, the Lorentz transformation law 
for the product of two spacetime 
fields depends on their spins.
\end{enumerate}
It is still an open question
how this twisted product and Lorentz transformations 
are seen in the effective field theory.
At least, if we restrict our attention to open string vertex 
operators only, there is another natural decomposition of the twist $\CF_1$,
which is closely related to the twisted Poincar\'e algebra 
on the Moyal-Weyl space as we will see in the next subsection.

Note also that there is another (and more conventional) 
treatment of a $B$-field background as a perturbation,
where the boundary action $S_B$ is considered as an interaction 
vertex operator and $e^{-S_B}$ is treated as perturbative insertion.
This corresponds to considering the $B$-field as a matter field in the effective field theory.
In that case, the quantization is the standard procedure, but the Poincar\'e symmetry is
explicitly broken by the background flux $B$.
It would be interesting to establish the equivalence to 
our treatment. (The situation is similar to the equivalence between 
commutative and noncommutative description of a D-brane worldvolume 
in Ref.\cite{Seiberg:1999vs}.)
The decomposition of the twist element here is suitable for that 
purpose, because 
the quantization scheme remains the standard one by $\CF_0$.
Thus, it would be possible to show that infinitely many perturbative 
insertions of $S_B$ reproduces the second twist $\CF_B$.
But we do not discuss this issue in this paper.

To summarize, from the decomposition of the twist element $\CF_1=\CF_B \CF_0$,
we obtain a description of a spacetime with a $B$-field background, 
such that the effect of background $B$-field is 
hidden in the twist element $\CF_B$,
while the other matter fields acquires a deformation 
caused by this twist element.
It seems that there is no other natural decomposition in the case of
 closed string vertex operators.
This description of spacetime 
is different from the description formulated on 
a commutative spacetime with a matter field $B_{\mu\nu}$ or 
from the description using 
a noncommutative space as in the effective theory of open string.
Although it is not clear yet how this new description translates into the effective field theory language, our method clearly demonstrates 
how a background field  (other than the metric) can be incorporated into a formulation of a stringy spacetime.

\subsection{Relation to field theory and twisted Poincar\'e symmetry}
\label{sec:field theory}

In this subsection, we study the relation between the Hopf algebra 
$\CH_{\CF_1}$ of string worldsheet theory and the corresponding Hopf algebra 
structure in the effective field theory on D-branes,
where the twisted Poincar\'e symmetry is realized in the field theory on the
Moyal-type noncommutative space \cite{Chaichian:2004za,Koch:2004ud,Aschieri:2005yw}.
As mentioned in section  1, one of the motivations of this work is
to understand the Hopf algebra symmetry and its twist in the full string theory.
Schematically, we want to establish the diagram
\begin{align}
\begin{array}{ccc}
\CH_{\CF_1} & \xrightarrow{\text{field theory}} & U_{\CF_M}(\CP') \\
\triangledown && \triangledown \\
\CA_{\CF_1} & \xrightarrow{\text{field theory}} & A_{\CF_M}.
\end{array}
\label{string-field diagram}
\end{align}
$\CH_{\CF_1}$ and $\CA_{\CF_1}$ are 
the Hopf algebra and module algebra 
in the string worldsheet theory, respectively.
On the r.h.s., $U_{\CF_M}(\CP')$ and $A_{\CF_M}$ are
 the corresponding Hopf algebra and module algebra structures in the effective field theory on the D-brane worldvolume, respectively.
$U_{\CF_M}(\CP')$ is the twisted Poincar'e symmetry 
with the Moyal twist $\CF_M$ and 
$A_{\CF_M}$ denotes the twisted module algebra with the Moyal product.
The structure of the objects on the r.h.s. 
 is studied in Refs.\cite{Chaichian:2004za,Koch:2004ud, Aschieri:2005yw}.
The structure of the objects on the l.h.s. is investigated 
in the first part of this paper.
The lower arrow in (\ref{string-field diagram}), 
i.e., the relation between OPE of vertex operators 
and the Moyal product is clarified in \cite{Seiberg:1999vs}.
In the following, we will complete diagram by showing 
the missing relation, i.e., the upper arrow in (\ref{string-field diagram})
between twisted Hopf algebra and the twisted Poincar\'e symmetry 
on the Moyal-Weyl space.

Let the worldsheet $\Sigma$ 
be the upper half plane, 
so that we can see the correspondence with the tree-level effective field theory.
Recall that the effective field theory on D-branes in the $B$-field background 
is obtained from the 
worldsheet theory by an appropriate moduli integral of the 
correlation functions of boundary vertex operators 
(see for example Ref.\cite{Seiberg:1999vs}).
There are two new ingredients when we move from correlation functions 
to the effective theory:
i) fixing the cyclic ordering of the insertions, and ii) zero slope limit.
Note that each cyclic ordering corresponds to a different region in the moduli space, 
and thus gives a different interaction term, 
e.g. $\int \Phi_1 \cdot(\Phi_2 * \Phi_3) \ne \int \Phi_1 \cdot(\Phi_3 * \Phi_2)$
for fields $\Phi_i$ on D-branes.
This is the source for the non-equivalence between planar and non-planar diagrams 
in the case of the noncommuative field theory.
It turns out that the ordering of insertions
 is the essential part to establish the above correspondence, 
however the zero slope limit becomes relevant
 when we consider an explicit form of the effective 
action.
Here we concentrate on the structure independent of a zero slope limit.

For our purpose here, it is appropriate to consider 
the twisted Hopf algebra $\CH_{\CF_1}$ with 
successive twists\footnote{Our argument here is restricted 
only to boundary vertex operators for open strings,
unless stated explicitly.
We will give a remark about closed string vertex 
operators at the end of this subsection.
} different from those
in section  \ref{sec:2step twist}.
 Therefore, we consider here the second decomposition, 
where the first twist is a quantization 
with respect to the open string metric, when restricted to boundary vertex operators,
and the second is its deformation.
This second twist turns out to contain the same information 
like the Moyal-twist on the D-brane and 
the universal R-matrix.

\paragraph{The second decomposition.}
In section \ref{sec:2step twist}, the successive twist 
is characterized by the decomposition of
the propagator (\ref{G_1}).
However, the decomposition of propagator is in fact arbitrary
and we can choose an appropriate decomposition 
depending on which physical property of the twisted 
Hopf algebra $\CH_{\CF_1}$ we want to see.
Here it is useful to decompose the propagator 
according to the symmetry of the tensor indices $\mu\nu$.
We denote the ``symmetric part" $G_S$ 
and the ``anti-symmetric part" $G_A$
of the propagator\footnote{Of course, the propagator 
$G_1^{\mu\nu}(z,w)$ is symmetric under the simultaneous exchanges
$\mu \leftrightarrow \nu$ and $z \leftrightarrow w$. 
This corresponds 
to the fact that the twisted module algebra with the product $*_{\CF_1}$
is commutative, and that the product in the path integral should be the time-ordered product.}, respectively.
For the upper half plane, the decomposition of the propagator is 
\be
G_1^{\mu\nu}(z,w)=G_S^{\mu\nu}(z,w)+G_A^{\mu\nu}(z,w), 
\label{SA decomposition}
\ee
where 
\bea
&&G_S^{\mu\nu}(z,w):=-{\alpha'}\left[ \eta^{\mu\nu}\ln |z-w| -\eta^{\mu\nu}\ln |z-\bar w|
+G^{\mu\nu}\ln |z-{\bar w}|^2 \right],  \nonumber\\ 
&&G_A^{\mu\nu}(z,w):=-{\alpha'} {\Theta^{\mu\nu}} \ln 
\frac{z-{\bar w}}{{\bar z}-w}~. 
\label{SA propagator}
\eea
If $z$ and $w$ are at the boundary, they reduce to the form
\be
G_S^{\mu\nu}(s,t)=-{\alpha'}G^{\mu\nu}\ln(s-t)^2 
\and G_A^{\mu\nu}(s,t)=\frac{i}{2} {\theta^{\mu\nu}} \epsilon (s-t),
\label{SA boundary}
\ee
where $\epsilon (t) $ is the sign function.

The following procedure is rather analogous to 
the one in section  \ref{sec:2step twist},
that is, 
we divide the twist into $\CF_1=\CF_A \CF_S$, and
we regard the first twist by the twist element
\begin{equation}
\CF_S=\exp\left(-{\int \!d^2z \!\!\int \!d^2w \,
G_S^{\mu\nu}(z,w)
\frac{\delta}{\delta X^{\mu}(z)}
\otimes\frac{\delta}{\delta X^{\nu}(w)}}\right),
\label{eq:F_S}
\end{equation}
as a quantization, 
and the second twist by the twist element
\begin{equation}
\CF_A=\exp\left(-{\int \!d^2z \!\!\int \!d^2w \,
G_A^{\mu\nu}(z,w)
\frac{\delta}{\delta X^{\mu}(z)}
\otimes\frac{\delta}{\delta X^{\nu}(w)}}\right),
\label{eq:F_A}
\end{equation}
as a deformation of the operator product, respectively.

\paragraph{First twist as quantization.}
The twisting by $\CF_S$ converts $\CH$ into
a twisted Hopf algebra $\CH_{\CF_S}$, which 
is isomorphic to $\hat{\CH}_S$ by elements of 
the form $\tilde{h}=\CN_S h \CN_S^{-1}$.
Correspondingly, we have a twisted module algebra $\CA_{\CF_S}$ with the 
product $*_{\CF_S}$, and the 
normal ordered module algebra $\hat{\CA}_S$.
We denote elements of $\hat{\CA}_S$ as
$\,^{\bullet}_{\bullet} F\,^{\bullet}_{\bullet} =\CN_S \act F$.
This defines a quantization scheme, but it turns out that 
it is natural only for boundary elements of $\CH$ and $\CA$.\footnote{
They are defined by the functionals and functional derivatives 
of the embedding $X^\mu (t)$ of the boundary, or equivalently, defined by inserting 
a delta function of the form $\int_{\p\Sigma} dt\delta^{(2)} (z-t)$ into $\CH$ and $\CA$.}

First, the Lorentz generator $L_{\mu\nu}$ in the Poincar\'e-Lie algebra acquire the twist and 
becomes non-primitive. 
This is the same situation considered in section \ref{sec:2step twist}.
However, in this case, there are other boundary elements in the classical Hopf algebra $\CH$, 
\be
L'_{\mu\nu} = \int_{\p\Sigma} \! dt \, G_{[\mu \rho}X^{\rho}(t) \frac{\delta}{\delta X^{\nu]}(t)}
\label{L'}
\ee
These $L'_{\mu\nu}$ and the translation generators $P_{\mu}$ (as boundary elements) constitute
another Poincar\'e-Lie algebra $\CP'$ when acting on boundary local functionals,
where the commutation relations are written 
with respect to the open string metric $G_{\mu\nu}$.
It is easy to show that the Hopf subalgebra $U(\CP')$ of $\CH$ is invariant 
under the twist $\CF_S$.
In particular, $L'_{\mu\nu}$ is primitive.
In terms of the normal ordered Hopf algebra, 
this means 
$\tilde{P_{\mu}}=P_{\mu}$ and $\tilde{L}'_{\mu\nu}=L'_{\mu\nu}$ 
as boundary elements in $\hat{\CH}_S$.
Therefore, $U(\CP')$ is considered to be a quantum Poincar\'e symmetry in this quantization scheme
when restricted on the boundary.

Consequently, boundary elements of the module algebra $\hat{\CA}_S$ are classified by 
the representation of $U(\CP')$ with a fixed momentum $k_\mu$.
A local functional $F[X]=e^{ik_\mu X^\mu(t)}$ defines
a tachyon vertex operator $\CN_S \act F= 
\,^{\bullet}_{\bullet} e^{ik\cdot X(t)}\,^{\bullet}_{\bullet}$ 
with momentum $k_\mu$. 
In general, a local boundary vertex operator $V_k (t)$ 
with momentum $k_\mu$ 
 consists of the worldsheet derivatives of $X$ and $e^{ik\cdot X}$,  
having the form 
\be
V_k (t)=\,^{\bullet}_{\bullet} P[\p X(t)] e^{ik\cdot X(t)} \,^{\bullet}_{\bullet}~~.
\label{boundary vertex}
\ee
Here, $P[\p X]$ is a polynomial in $\p_t^l X^\mu (t)$ and $\p_n^l X^\mu (t) \ (l=1,2,\cdots)$.
The functional form of $P[\p X]$ determines the 
representation of $U(\CP')$.
Since the action of $h \in U(\CP')$ is given by $h \act V_k (t) = \,^{\bullet}_{\bullet} 
h \act P[\p X(t)] e^{ik\cdot X(t)} \,^{\bullet}_{\bullet}$, 
this representation is the same as that of a classical functional.
For example, $\,^{\bullet}_{\bullet} \p_t X^{\mu} e^{ik\cdot X(t)} \,^{\bullet}_{\bullet}$
transforms as a $1$-form and the lowering and the raising of tensor indices are 
taken with respect to $G_{\mu\nu}$.
Note that above boundary vertex operators $V_k (t)$ are 
equivalent to the vertex operators used in Ref.\cite{Seiberg:1999vs}.
Their anomalous dimension and the on-shell condition are determined 
with respect to the open string metric $G_{\mu\nu}$.

\paragraph{Second twist.}
Twisting by $\CF_A$, we obtain  the twisted Hopf algebra $(\hat{\CH}_S)_{\CF_A}$, 
which is equivalent to the full twisted Hopf algebra $\CH_{\CF_1}$ 
as in the case discussed in section \ref{sec:2step twist}.
Correspondingly, we obtain the twisted module algebra $\CA_{\CF_1} \simeq (\hat{\CA}_S)_{\CF_A}$.
The twisted product $*_{\CF_1}$ in $\CA_{\CF_1}$ 
is seen as a deformation of the product in $\hat{\CA}_S$:
\be
\langle\,^{\circ}_{\circ} F \,^{\circ}_{\circ} \,^{\circ}_{\circ} G \,^{\circ}_{\circ}
\,\rangle_1
= \tau( F*_{\CF_1} G )
= \langle\,^{\bullet}_{\bullet} F\,^{\bullet}_{\bullet}
*_{\CF_A} \,^{\bullet}_{\bullet}G\,^{\bullet}_{\bullet}\,\rangle_S.
\ee
This is valid for arbitrary functionals $F,G \in \CA$, 
including bulk local functionals corresponding to the closed string vertex operators.

If we consider only the boundary vertex operators, 
there is a great simplification 
in the structure of the second twist.
This is because the boundary-boundary propagator 
$G_A^{\mu\nu}(s,t)$ in (\ref{SA boundary}) is topological, 
i.e., it depends only on the sign of $s-t$.
The twist element acting on the boundary vertex operators has the form
\bea
\CF_A &=& \exp\left\{-\frac{i}{2} {\theta^{\mu\nu}}
\int \! ds \!\! \int \! dt \,\epsilon (s-t)
\frac{\delta}{\delta X^\mu(s)}
\otimes \frac{\delta}{\delta X^\nu(t)}\right\}~~.
\label{F_A}
\eea
First, worldsheet derivatives of $X$ do not feel this deformation:
e.g.,
$\p_a X^{\mu}(s) *_{\CF_A} X^\nu(t)= \p_a X^{\mu}(s) X^\nu(t)$
holds for $s\ne t$.
Therefore, for any correlation function of boundary vertex operators of the form 
(\ref{boundary vertex}),
the product $*_{\CF_A}$ is only sensitive to the tachyon part, and 
we have the relation
\be
\langle\,V_{k_1}(t_1) *_{\CF_A} \cdots *_{\CF_A} V_{k_n}(t_n) \,\rangle_S
= e^{-\frac{i}{2}\sum_{i>j} k_{i\mu} \theta^{\mu\nu} k_{j\nu} 
\epsilon (t_i-t_j) }
\langle\,V_{k_1}(t_1) \cdots V_{k_n}(t_n) \,\rangle_S ~~
\label{correlation function}
\ee
Next, by fixing the order of insertion points $t_1 > t_2 > \cdots > t_n$, 
the extra phase factor in the r.h.s. above becomes independent of the precise locations,
and gives the factor of the Moyal product.
As explained in Ref.\cite{Seiberg:1999vs}, this
 enables us to take the following prescription:
{\it replace ordinary multiplication in the effective field theory written in the 
open string metric by the Moyal product.}
We emphasize here that this works independently of the kind of the field (representation of $U(\CP')$) 
on D-branes,  contrary to the case in section \ref{sec:2step twist}.
On the contrary, if $*_{\CF_A}$ would be sensitive to the derivatives of $X$, 
then the definition of the product for scalar fields and vector fields,
such as $\phi * \phi$ and $\phi * A_\mu$, would be different.
In the Hopf algebra language, this means that 
with a fixed ordering the twist element (\ref{F_A}) acts as 
a Moyal-twist  
\be
\CF_M = e^{\frac{i}{2} {\theta^{\mu\nu}} P_\mu \otimes P_\nu}~~, 
\ee
given in the literature 
Refs.\cite{Oeckl:2000eg,Watts:2000mq,Chaichian:2004za,
Koch:2004ud,Aschieri:2005yw}.
Here, the integrals of functional derivatives in (\ref{F_A}) 
are replaced by its zero mode part, the translation generators $P_\mu \in U(\CP')$.

The Poincar\'e symmetry is also simplified. 
The coproduct of the Lorentz generators $L'_{\mu\nu} \in U(\CP')$ 
acting on boundary vertex operators is twisted by $\CF_A$ as
\be
\Delta_{\CF_A} (L'_{\mu\nu}) = \Delta (L'_{\mu\nu})
+\frac{1}{2} {\theta^{\alpha \beta }}
\int \! ds \!\! \int \! dt \,\epsilon (s-t) G_{\alpha[\mu }
\frac{\delta}{\delta X^{\nu]}(s)}
\otimes \frac{\delta}{\delta X^\beta (t)}
\ee 
but it reduces by fixing the ordering $s>t$ to 
\be
\Delta_{\CF_M} (L'_{\mu\nu}) =\Delta (L'_{\mu\nu})
+\frac{1}{2} {\theta^{\alpha \beta }}
\left\{
 G_{\alpha[\mu }P_{\nu]} \otimes P_\beta
 + P_\alpha  \otimes G_{\beta [\mu } P_{\nu]}
\right\},
\ee
This gives the twisted Poincar\'e-Hopf algebra $U_{\CF_M} (\CP')$ structure
found in Refs.\cite{Chaichian:2004za,Koch:2004ud,Aschieri:2005yw}.
Note that since the twist element 
itself now belongs to $U(\CP') \otimes U(\CP')$,
the twisting is closed in $U(\CP')$, 
contrary to the case in section \ref{sec:2step twist}.
To summarize, the twisted Poincar\'e symmetry on the Moyal-Weyl noncommutative space 
is derived from string worldsheet theory in a $B$-field background.
We give two further remarks.

\paragraph{Relation between $\CF_A$ and $\CF_M$.}

Before fixing the ordering of insertions, we have
\bea
&&\CF_A^{-1} \act \left(V_{k_1}(t_1) \otimes V_{k_2}(t_2)\right) \nonumber \\
&&= \exp\left\{-\frac{i}{2} {\theta^{\mu\nu}}
\int \! ds_1 \!\! \int \! ds_2 \,\epsilon (s_1-s_2) k_{1\mu}k_{2\nu} 
\delta (s_1-t_1)\delta (s_2-t_2)\right\}
\left(V_{k_1}(t_1) \otimes V_{k_2}(t_2)\right) \nonumber \\
&&= e^{-\frac{i}{2} \epsilon (t_1-t_2) {\theta^{\mu\nu}}k_{1\mu}k_{2\nu} }
\left(V_{k_1}(t_1) \otimes V_{k_2}(t_2)\right) \nonumber \\
&&=\left[
\theta (t_1-t_2) e^{-\frac{i}{2}{\theta^{\mu\nu}}k_{1\mu}k_{2\nu}}
+\theta (t_2-t_1) e^{\frac{i}{2}{\theta^{\mu\nu}}k_{1\mu}k_{2\nu}}
\right]
\left(V_{k_1}(t_1) \otimes V_{k_2}(t_2)\right) \nonumber \\
&&=\left[
\theta (t_1-t_2) \CF_M^{-1} +\theta (t_2-t_1) \CF_M
\right] \act 
\left(V_{k_1}(t_1) \otimes V_{k_2}(t_2)\right),
\eea
using the fact that $\CF_A$ acts only on the tachyon part, and 
$\epsilon(t)=\theta (t)-\theta (-t)$.
Therefore, on the tensor product $V_{k_1}(t_1) \otimes V_{k_2}(t_2)$ of boundary vertex operators, we get a relation 
\bea
\CF_A^{-1} &=& 
\theta (t_1-t_2) \CF_M^{-1} +\theta (t_2-t_1) \CF_M \nonumber\\
&=& \CF_M^{-1} \left[\theta (t_1-t_2)  +\theta (t_2-t_1) \CR_M^{-1}
\right] .
\label{F_A F_M relation}
\eea
Here $\CR_M:=\CF_{M21}\CF_M^{-1}$ is the universal R-matrix for the Moyal deformation, 
and its inverse is 
$\CR_M^{-1}=\CF_M^2 = e^{i{\theta^{\mu\nu}}P_\mu \otimes P_\nu}$. 
In general, the universal R-matrix 
is given for an {\it almost} cocommutative Hopf algebra, 
in the sense that the coproduct is cocommutative up to conjugation by $\CR$.
\be
\Delta^{op} (h) =\CR \Delta (h) \CR^{-1}.
\ee
Correspondingly, the twisted product $f*g$ in the module algebra $\CA_{\CF}$ is related 
to its opposite product $g*f$ as
\bea
g* f &=& m \circ \CF^{-1} \act (g\otimes f) \nonumber \\
&=& m \circ \CF_{21}^{-1} \act (f\otimes g) \nonumber \\
&=& m \circ \CF^{-1} \CR^{-1} \act (f\otimes g)~.
\eea
Hence, $*$ is almost commutative up to the insertion of $\CR^{-1}$.

In our case, the twist $\CF_A$ in the string theory is strictly commutative,
and therefore the $R$-matrix is trivial, 
$\CR_A =1\otimes 1$.
It corresponds to the time ordered product in the operator formulation.
However, as shown in (\ref{F_A F_M relation}) the product of functionals is 
rewritten by the operator product with fixed ordering, where
each operator product defined by $\CF_M$ is noncommutative but almost commutative.
In this way the noncommutative product in the effective field theory is
derived from the twisted but commutative product in the worldsheet theory. 

\paragraph{Remark on closed strings}

In the above discussion, the insertion points of the vertex operators 
are restricted to the boundary of the world sheet.
Owing to this restriction, the second twist $\CF_A$ 
reduces to the Moyal-twist, and 
we obtain the twisted Poincar\'e-Hopf algebra $U_{\CF_M} (\CP')$.
On the other hand, if we consider the 
correlation function of closed string, or 
couplings of D-branes to the closed string, the insertion points are
not always at the boundary and
it is inevitable to use the bulk-bulk or the bulk-boundary propagator.
Therefore, $U_{\CF_M} (\CP')$ is not sufficient 
to describe this situation, rather we should work within 
a full diffeomorphism $\CH$ and twisting by $\CF_1$.
This is one reason why 
the simple generalization 
of the twisted Poincar\'e symmetry to 
a diffeomorphism \cite{Aschieri:2005yw,Aschieri:2005zs} 
based on the Moyal-twist $\CF_A$
is not sufficient to describe a symmetry in a brane induced gravity, 
as pointed out in Ref.\cite{AlvarezGaume:2006bn}.
It is an interesting question 
how the twisted diffeomorphism in the 
effective field theory has to be modified
in order to recover the string theory result.

\section{Conclusion and discussion}


Considering the Hopf algebra structure of the symmetry in 
string theory, the quantization 
of the string can be considered as
the Drinfeld twist. 
For the case of the trivial background, i.e., in the Minkowski
target space with vanishing $B$ field, we have shown in Ref.\cite{AMW} 
that the path integral quantization of the string worldsheet theory
is reformulated as the Drinfeld twist of the Hopf algebra $\CH$
and the module algebra $\CA$ of functionals on the worldsheet, with twist element $\CF_0$.
The quantized algebra of the vertex operators $\hat{\CA}$ appeared via 
the coboundary relation (\ref{coboundaryrelation})
which showed that the twist characterizes also 
the normal ordering $\CN_0$, a property which is rather implicit in the path integral.

In the present paper, we applied the same formalism
to the case with non-vanishing 
$B$-field background. The quantization is characterized by 
the twist element $\CF_1$.  
In the twist quantization, we can also obtain the natural choice
of the normal ordering.  We have discussed the mechanism
to obtain the information on the normal ordering and 
have shown that the coboundary relation
for the twist element $\CF_1$ yields that the natural normal ordering
in this approach is given by $\CN_1$.
The normal ordering
conditions obtained in this way is most natural from the point of view
of the Hopf algebra as the symmetry of the theory.

One of the advantages of the twist quantization is that there is
a direct relation between quantization and symmetry 
via the Hopf algebra.
For example, the Poincar\'e symmetry $U(\CP)$ is also 
twisted by $\CF_1$ in the twist quantization.
In order to compare the twist quantization with known formulations,
we have decomposed the twist into two successive twists.
We have discussed two types of decomposition.

In the first decomposition, $\CF_1=\CF_B \CF_0$, 
 the first twist by $\CF_0$ is the standard quantization of string 
with $B=0$ and the second twist by
$\CF_B$ is a twist of the operator product of vertex operators by the $B$-field. 
The resulting structure is the twisted Poincar\'e symmetry,
which indicates the new description of spacetime, 
with the $B$-field as a twist rather than a background field.
The description here can be considered as an example, for incorporating 
a closed string state into a more general kind 
of geometry (stringy geometry).
This decomposition is also natural 
if one wants to compare 
with the perturbative calculation of the effect of the 
$B$-field background.

The second decomposition, $\CF_1=\CF_A \CF_S$, 
is meaningful only for open strings, 
where the first twist $\CF_S$ is interpreted as a quantization 
with respect to the open string metric, 
and the second twist $\CF_A$ reduces to the Moyal twist $\CF_M$ 
equipped with the universal R-matrix.
This explains clearly the relation between the 
twisted Poincar\'e symmetries 
in the string worldsheet theory and 
in the low energy effective theory on D-branes.
From the viewpoint of noncommutative field theory,
this second decomposition gives an alternative way to 
Refs.\cite{Aschieri:2005yw,Aschieri:2005zs},
for generalizing the noncommutativity of the Moyal plane, 
where in our case the (open string) metric $G_{\mu\nu}$ 
is also incorporated in the twist $\CF_1$.
Moreover, the concept of the twisted diffeomorphism 
in Refs.\cite{Aschieri:2005yw,Aschieri:2005zs}
is realized not at the level of the noncommutative 
field theory, but at the level of the worldsheet theory. 
As stated in the end of section \ref{sec:field theory}, 
the true twisted diffeomorphism on the Moyal plane should be different from that in 
Refs.\cite{Aschieri:2005yw,Aschieri:2005zs} as a low energy effective theory on D-branes.

The decomposition is rather arbitrary and possible as far as 
the second twist satisfies the cocycle condition 
under the first twist, the condition 
corresponding to (\ref{2nd cocycle condition}).
Therefore, the method proposed in this paper would be applicable 
to more non-trivial backgrounds, 
such as non-constant metric or $B$-field.
The corresponding perturbative calculations in these 
cases have been done in Refs.
\cite{Cornalba:2001sm,Ho:2001qk,Herbst:2001ai,Guttenberg:2007ce}.
It is an interesting question whether their results can be read off 
as an analogue of the first decomposition, $\CF_1=\CF_0\CF_B$
 of some (possibly non-abelian) twist element.
This would help to get a better understanding of the structure 
of the stringy geometry.

As already remarked in Ref.\cite{AlvarezGaume:2006bn}, 
the string field theory is 
a formulation which can, in principle, represent the space-time 
symmetry as a kind of 
gauge symmetry.
The constant $B$-field background in such a formulation 
has been considered in Refs.\cite{Kawano:1999fw,Kawano:2000uq}.
There, a product of string fields (string vertex) acquires a 
modification by a phase factor, 
which is similar to the twist element $\CF_A$ here.
Thus, the Hopf algebra structure and the method presented here
 would also be helpful to understand 
the structure of the gauge symmetry 
in the string field theory.

\acknowledgments{
The authors would like to thank
B.~Durhuus, K.~Ohta, H.~Ishikawa and 
for useful discussions. We also would like to acknowledge 
U. Carow-Watamura for discussions and valuable comments.
This work is supported in part by the Nishina Memorial Foundation (T.A.)
and by Grant-in-Aid for Scientific Research from the Ministry of Education, Culture, Sports, Science and Technology, Japan, No. 19540257(S.W.).}

\end{document}